\documentclass[prl,twocolumn,superscriptaddress]{revtex4-1}
\usepackage{graphicx}
\usepackage{amsmath}
\usepackage{dcolumn}
\usepackage{color}
\usepackage{epstopdf}
\usepackage[caption=false]{subfig}

\begin{document}  

\title{Plasmoids formation during simulations of coaxial helicity injection in the National Spherical Torus Experiment}
\author{F. Ebrahimi}
\affiliation{Department of Astrophysical Sciences, Princeton University, Princeton, NJ 08543}
\author{R. Raman}  
\affiliation{University of Washington, Seattle, WA 98195}

\date{\today}
  
\begin{abstract}
  Formation of an elongated Sweet-Parker current sheet and a transition to plasmoid instability has for the first time been predicted by simulations in a large-scale toroidal fusion plasma in the absence of any pre-existing instability.  Plasmoid instability is demonstrated through resistive MHD simulations of transient Coaxial Helicity Injection (CHI) experiments in the National Spherical Torus Experiment  (NSTX). 
 Consistent with the theory,  fundamental
 characteristics of the plasmoid instability, including fast reconnection rate, have been observed in these realistic simulations.  Motivated by the simulations, experimental camera images have been revisited and suggest the existence of reconnecting plasmoids in NSTX. Global, system-size plasmoid formation observed here should also have strong implications for astrophysical reconnection, such as rapid eruptive solar events.
\end{abstract}
\maketitle
Magnetic reconnection, the rearrangement of magnetic field
topology of plasmas, energizes many processes in nature, such as solar and stellar flares, magnetospheres, coronas of accretion disks and astrophysical jets. Magnetic reconnection has also been demonstrated to be critical in the nonlinear dynamics of many processes in toroidal fusion plasmas, such as sawtooth oscillations, plasma disruption, plasma relaxation and magnetic self-organization (see for example ~\cite{zweibel09}). Sweet-Parker (S-P) steady-state reconnection~\cite{sweet58,parker57} as a local approach and spontaneous magnetic reconnection as the result of tearing
fluctuations~\cite{fkr,coppi76} have been the two primary physical models used to explain these processes. 
Although these models with classical dissipative rates (in which the reconnection rate scales with Lundquist number with a negative power) have not been able to explain fast reconnection rates in astrophysical systems (with large Lundquist numbers) or the fast time scales of sawtooth crashes, they have explained major reconnection signatures such as strong outflows or growth phase and current and momentum relaxation~\cite{choiprl,ebrahimi2007}. Sweet-Parker type forced magnetic reconnection in laboratory plasmas~\cite{ji98,yamada2000} and 
spontaneous reconnection (non-axisymmetric tearing instabilities) in toroidal fusion plasma has been extensively but separately studied. However, the Sweet-Parker type forced reconnection and spontaneous reconnection can be related through plasmoid instability.

In large-scale collisional plasmas with sufficiently high Lundquist numbers, elongated S-P current layers become tearing unstable and break up into multiple islands or plasmoids~\cite{biskamp86,shibata2001}. According to linear theory~\cite{loureiro07}, the instability has a super-Alfvenic growth rate increasing with $S$ as $S^{1/4} V_A /L$  and the number of plasmoids  increasing as $S^{3/8}$, where $S$ is the dimensionless Lundquist number $S = L V_A/\eta$, $V_A$ is the Alfven velocity based on the reconnecting 
magnetic field,  $L$ is the current sheet length, and $\eta$ is the magnetic diffusivity. In recent years, several authors have developed systematic theoretical and numerical MHD and fully kinetic studies of plasmoid instability in slab geometries (with equilibria such as Harris sheet type models)~\cite{bhattacharjee09,samtaney09,huang2010,lapenta08,daugthon09}. 
It has been found that plasmoids allow a fast
reconnection rate in the resistive MHD model, nearly independent of Lundquist number at high $S$. These
recent developments have taken the reconnection model
of resistive MHD beyond the S-P model and made the
MHD model again relevant for fast reconnection. However, effects beyond MHD may also contribute to fast reconnection as the current sheet width ($\delta_{\mathrm{sp}}$) becomes smaller than the
two-fluid or kinetic scales~\cite{cassak2005,ji2011}. Secondary islands (plasmoids) have also been seen in reduced MHD simulations during the nonlinear evolution of the tearing instability in slab geometry~\cite{loureiro05} and during the nonlinear growth of an internal kink mode in cylindrical geometry~\cite{yu14,gunter15}. In these studies it was concluded that for realistic fusion relevant parameters, it is the inclusion of two-fluid effects, not the plasmoid instability, that may lead to the fast sawtooth reconnection~\cite{gunter15}. Despite the numerous observational evidence of plasmoid-like structures in the Earth's magnetosphere and the solar atmosphere~\cite{shibata2001,kliem2010}, this instability has not been reported to exist in a laboratory plasma or a fusion device.
\begin{figure*}
\centering
\setlength\fboxsep{0pt}
\setlength\fboxrule{0.25pt}
\fbox{\includegraphics[width=6.2in]{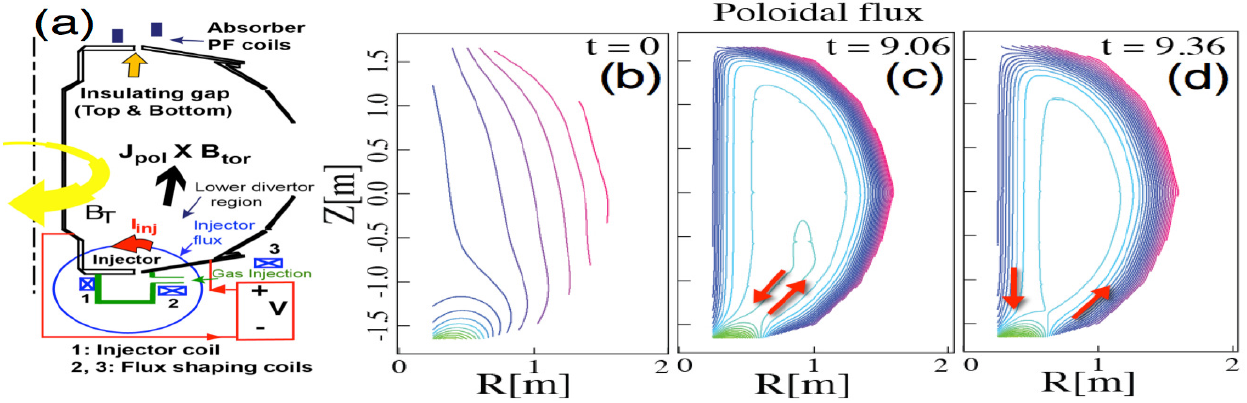}}
\caption{(a) Line drawing showing the main components in NSTX
required for plasma start-­‐up using CHI. The initial poloidal
field, the injector-­‐flux (shown by the blue ellipse), connecting the inner and outer divertor plates
in the injector region is produced using the lower divertor
coils (shown with numbers 1,2 and 3). In the presence of an 
external toroidal field ($B_T$), and after gas injection, voltage (V)
is applied to the divertor plates at $t=6$ms
(as in the simulations) to initiate the discharge~\cite{raman2006}. The path of $I_{\mathrm{inj}}$ from the voltage source to the outer divertor plate, then through the upper part of the injector flux, and back to the voltage source, is shown by the three red arrows. (b)-(d) Poloidal 
flux evolution during simulations with one
plasmoid.
After the discharge fills the vessel (c),
the voltage is rapidly reduced to zero at $t=9$ms.
This induces reconnection in the injector region to form closed flux surfaces (d). The arrows show oppositely directed 
field lines in the injector region where the
S-­‐P current sheet forms (c).}
 \label{fig:fig1} 
\end{figure*} 

In this letter, we examine the plasmoid formation instability in a large toroidal fusion plasma experiment using resistive MHD simulations. The novel characteristic of these simulations is that they have been performed in a realistic experimental geometry that includes currents driven in the external toroidal and poloidal field coils. It is demonstrated that during transient Coaxial Helicity Injection (CHI) discharges at high Lundquist number, the elongated current sheet formed through a Sweet-Parker forced reconnection process breaks up, and transition to a spontaneous reconnection (plasmoid instability) occurs. The transition to plasmoid instability is identified through: 1) the break up the elongated current sheet, 2) the increasing number of plasmoids with Lundquist number, and 3) the reconnection rate, as it becomes nearly independent of $S$. Plasmoid formation has importance for the fast flux closure observed during the experiments, because the simulations predict that the reconnection rate in the presence of plasmoids is faster than the S-P reconnection rate. Motivated by these simulations, old experimental camera images have been re-examined.  Interestingly, these images from the large NSTX Spherical Torus (ST) experiment (major radius=0.86m, minor radius=0.65m) suggest the existence of reconnecting plasmoids.

In this paper, the formation of an elongated current sheet and the transition to plasmoid instability is shown to occur in the \textit{absence of any pre-existing instability}, i. e. tearing or kink instability. This transition is examined during transient CHI, a form of electrostatic helicity injection and non-inductive current drive. Transient CHI which is a leading candidate for plasma start-up and current formation in NSTX and NSTX-U, has generated toroidal current on closed flux surfaces without the use of the conventional central solenoid~\cite{raman2006}. Elimination of the central solenoid is necessary for low aspect ratio ST based reactor. In CHI, as shown in Fig.~\ref{fig:fig1}, by driving current along the open field lines (the injector current $I_{\mathrm{inj}}$), helicity is injected
through the linkage of toroidal flux with the poloidal injector flux. Plasma and open field lines (the magnetic bubble) expand into the vessel if the injector
current exceeds a threshold value (for details see~\cite{raman2006},~\cite{jarboe89}). Understanding the dynamics and the mechanism of closed flux surface formation during transient CHI is of great importance for the extrapolation of the concept to larger devices and is being studied for the NSTX configuration. In~\cite{ebrahimi2014} using a systematic approach using MHD simulations, the fundamental reconnection mechanism that leads to the generation of closed flux surfaces was explained. As also shown through the evolution of poloidal flux surfaces in Fig.~\ref{fig:fig1}(b)-(d), it was found that closed flux surfaces expand in the NSTX global domain through a local S-P type reconnection~\cite{ebrahimi2013,ebrahimi2014}. Motivated by these simulations of forced magnetic reconnection (a universal concept similar to the solar corona's reconnection), it was recognized that the CHI discharges, from the large NSTX device, may also provide a rich platform for investigating further fundamental reconnection physics, in particular whether there is a transition to a spontaneous instability. %

\begin{figure}
\includegraphics[]{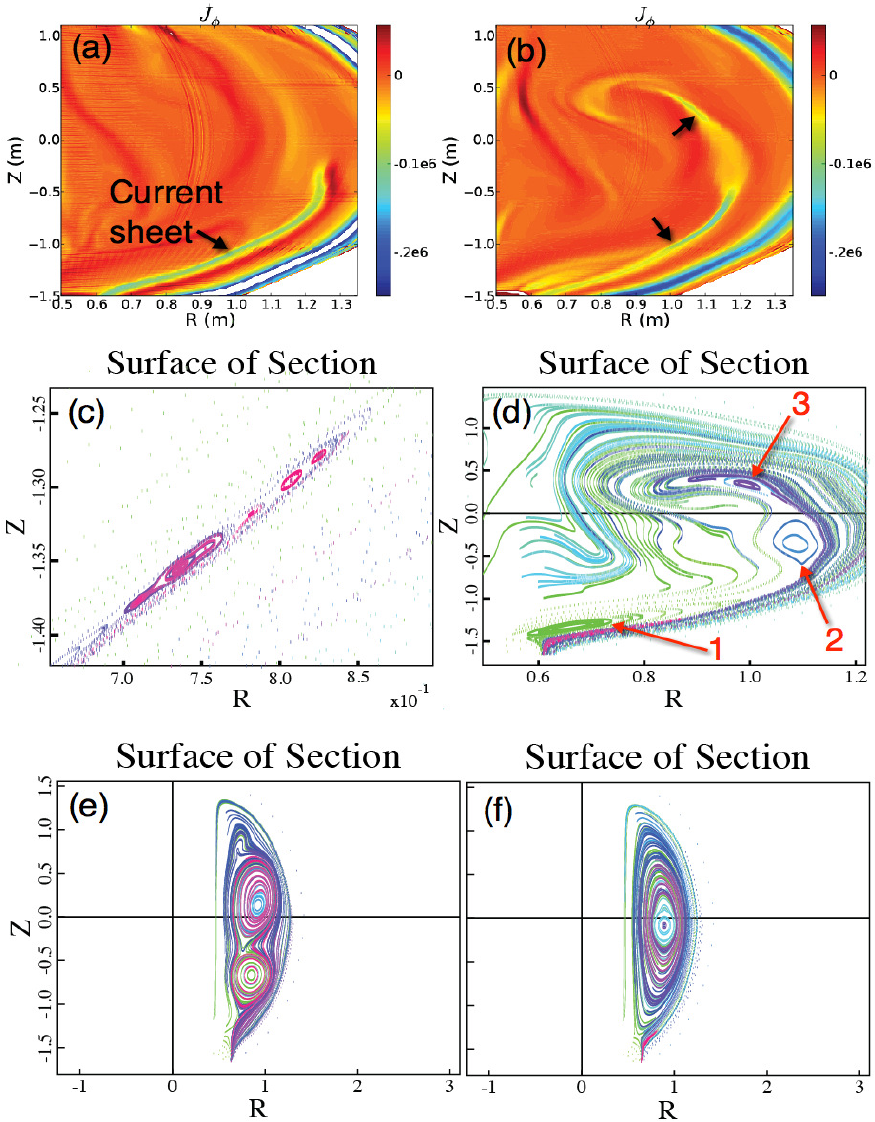}
\caption{(a) and (b) Poloidal R-Z cut of toroidal currents during magnetic reconnection and (c) and (d) Poincare plots at two times.  (a),(c) An unstable elongated current sheet formed (shown in yellow-green $L\approx$ 1.4m) and multiple small scale transient  plasmoids formed at the same time (t= 9.044ms). (b) and (d) A large scale breaking of current sheet to form three larger scale islands is clearly seen, current sheets between islands is S-P like (t= 9.174ms). (e),(f) Poincare plots during the evolution of large-scale plasmoids (system-size) into one plasmoid at, (e) t= 9.645ms (f) t=10.2ms, $S$=39000. The points in the Poincare plots are the intersections of a field line with a poloidal plane, as the field line is followed around the torus.  
}
  \label{fig:fig2} 
\end{figure}

To explore the transition to plasmoid instability, we have performed resistive MHD simulations of transient CHI in the NSTX using the NIMROD code~\cite{sovinec04}. To elucidate the underlying physics, using the NSTX vessel geometry, resistive MHD simulations for a zero pressure model are performed. In these simulations, we use constant poloidal field coil currents to generate the injector flux (fixed boundary flux simulations) shown in Fig.~\ref{fig:fig1}(b). The distance between the injector 
flux footprints is important
for driving an effective magnetic reconnection process and achieving maximum flux closure~\cite{ebrahimi2014}.  Here, to access the plasmoid instability regime, $S$ is increased through decreasing the magnetic diffusivity ($\eta$) or by increasing injector voltage using narrow flux footprints. The helicity injection model, boundary condition and geometry are the same as in the earlier papers~\cite{ebrahimi2013,hooper2013}. A uniform number density of $4 \times 10^{18} \mathrm{m}^{-3}$ for a deuterium plasma is used. Axisymmetric ($n$=0) 
simulations with poloidal grid 45 $\times$ 90 sixth to eighth
order finite elements are performed 
 in the geometry of the experiment. 
To obtain $S$ scaling, we used magnetic diffusivitites in the range of $2.5-20 m^2/s$. The kinematic viscosities are chosen to give a Pm =7.5 (Prandtl number = $\nu/\eta$) for all values of magnetic diffusivitites used here. A simplified waveform of injector voltage with a constant voltage is applied at 6ms and
turned off at 9ms. After this time, we observe the reconnection process as the field lines are brought together by a radial force.

It is first confirmed by these simulations that at sufficiently high $S$ (larger than 100), the oppositely directed field lines [shown in~Fig.~\ref{fig:fig1}(c)] in
the injector region have sufficient time to reconnect (before dissipating), leading to the
formation of closed flux surfaces~[Fig.~\ref{fig:fig1}(d)]. The formation of closed flux surfaces begins via a formation of a S-P current sheet in the injection region. In this case, the current sheet is stable and X point formation near the injection region gives rise to flux surface closure. As $S$ is increased (i.e $L/\delta_{\mathrm{sp}}$ is also increased), a transition from a stable current sheet to an MHD unstable one occurs. The current sheet formed in our simulation can extend to about $L$=2.5-3m, to the extent of the vertical elongation of the NSTX vessel. In this process, the elongated S-P current sheet becomes unstable to tearing instability and plasmoids, accompanied 
by multiple X points, are formed spontaneously. Poloidal R-Z cut of toroidal current densities and the Poincare plots during magnetic reconnection at two times for a plasmoid unstable case at $S$=39000 are shown in Fig.~\ref{fig:fig2}.  As the field lines reconnect at t= 9.044ms multiple small-scale islands (plasmoids) are formed around Z$\sim$ -1.4m to -1.25m as it is clearly seen from the Poincare plot in Fig~\ref{fig:fig2} (c). This is because at this time the elongated current sheet already falls into the 
plasmoid unstable regime
($L/\delta_{sp} \approx 70$). The process is very dynamical. Interestingly, these small plasmoids are transient and gradually merge as more poloidal flux reconnects. As magnetic reconnection continues and more field lines reconnect, the current sheet also continues to elongate. In this process, as the small transient plasmoids (Fig~\ref{fig:fig2} (c)) merge and form a single island at around R=0.6-0.8m and Z = -1.5 to -1.2, shown in Fig.~\ref{fig:fig2}(d), two new larger plasmoids at the upper part of the device are formed (Z$\sim$ -0.5m and Z $\sim$ 0.3m).  The current sheet as shown in Fig.~\ref{fig:fig2}(b) has broken up and three new islands form (Fig.~\ref{fig:fig2}(d)).  These larger scale islands that spread along the vertical direction are more persistent. However, even these large islands go through a dynamical process and move around and ultimately they do 
merge. Figures~\ref{fig:fig2}(d)-(f) show the dynamics of these large scale plasmoids as they merge in time. As seen from these Poincare plots,  three islands (Fig.~\ref{fig:fig2}d)  merge to two islands (Fig.~\ref{fig:fig2}e), two islands merge to one and form closed flux surfaces (Fig.~\ref{fig:fig2}f). 
\begin{figure}
\includegraphics[]{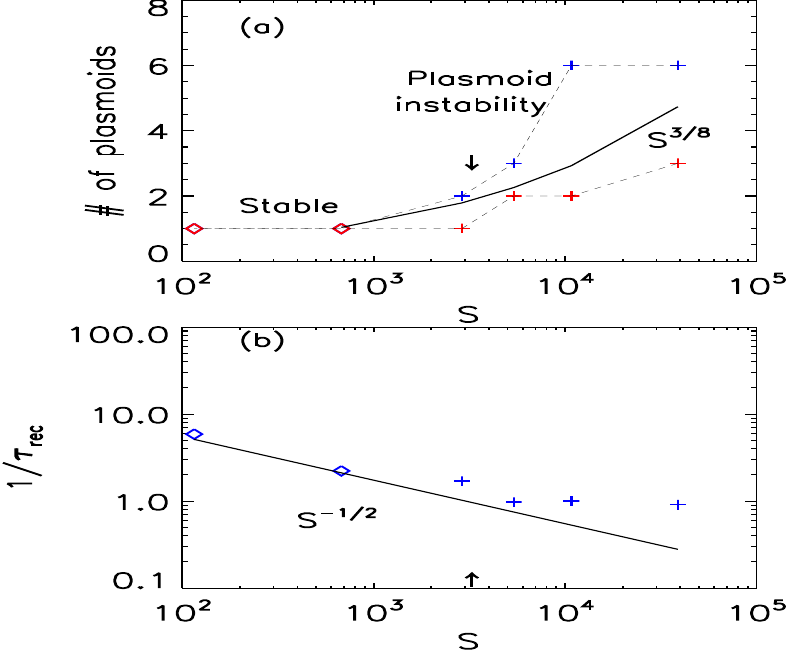} 
\caption{(a) Number of plasmoids vs. $S$. Blue: small sized transient plasmoids during the early phase of discharge (Fig.~\ref{fig:fig2}); Red: large scale and persistent plasmoids during the later phase of discharge. The solid line is the linear theoretical $S$ scaling. (b) The reconnection rate, $1/\tau_{rec}$ vs. $S$. The transition to plasmoid instability is shown at $S \sim$ 3000. The solid line is the S-P scaling.}
 \label{fig:fig3} 
\end{figure}

Through Poincare plots, we have identified the number of plasmoids generated during magnetic reconnection process. The number of plasmoids generated during magnetic reconnection as $S$ increases are shown in Fig.~\ref{fig:fig3}(a). If the current sheet is stable, only a single X point forms (S-P type reconnection) as shown in Fig.~\ref{fig:fig3} for $S<10^3$. At approximately $S$= 3000, where the transition to a plasmoid instability occurs, multiple X points form as shown in Fig.~\ref{fig:fig2}. The number of small transient plasmoids shown in blue is an increasing function of $S$. The number of more large scale and persistent plasmoids also increase with $S$ (shown in red). We have also calculated the reconnection time ($\tau_{\mathrm{rec}}$) defined here as the time from when the X point is formed until the formation of the largest closed flux volume.
Figure~\ref{fig:fig3}(b) shows the calculated reconnection rate (1/$\tau_{\mathrm{rec}}$) for different values of $S$ as the dynamics goes through a transition to an instability. 
As it is seen, when the current sheet is stable (S-P type reconnection), the reconnection rate follows the scaling of $S^{-1/2}$ (this regime was studied in~\cite{ebrahimi2013}). However, as the transition to plasmoid instability occurs, the reconnection rate becomes nearly independent of $S$, consistent with earlier studies in slab geometries~\cite{bhattacharjee09,loureiro2012}.
\begin{figure}
\includegraphics[]{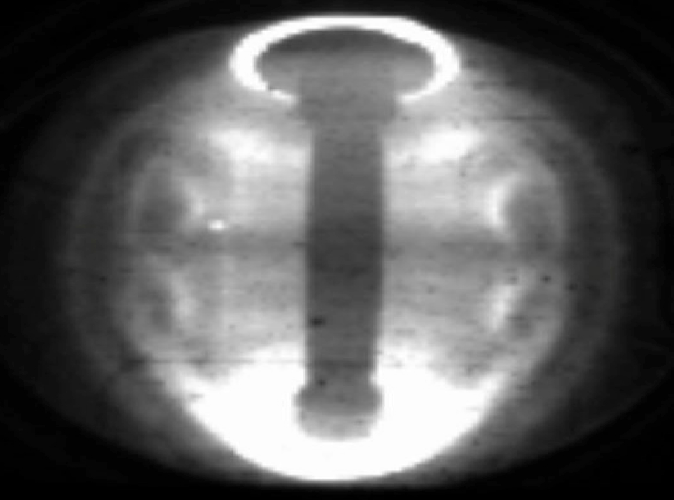}
\caption{Camera image during CHI discharge in NSTX shows the formation of two discrete plasma bubble plasmoids.}
 \label{fig:fig4} 
\end{figure}

Motivated by the MHD simulations presented here, which show
the formation of micro plasmoids
that tend to merge into a much larger plasma,
we have re-examined the very early phase of some NSTX discharges
to see if there is any experimental evidence for the formation of these plasmoids.
These measurements were carried out using a fast-framing camera
that viewed the entire NSTX vessel in a fisheye mode.
Although the camera imaging parameters were not optimized for capturing these early, highly dynamic features that are present during the early phase of the discharge,  some of the images suggest the formation of plasmoids, an example of which is shown in Fig.~\ref{fig:fig4}. The formation and the  vertical expansion of toroidally symmetric and plasmoid-like structures are also seen in the movie provided~\cite{movie}.  Although the camera images are encouraging,  more detailed measurements are needed in which these images are obtained at higher resolution, and from additional toroidal locations. In addition,  measurement of the plasma parameters at the reconnection layer itself are needed to determine the experimental parameter range over which these plasmoid-like features are observed.

\begin{figure}
\includegraphics[]{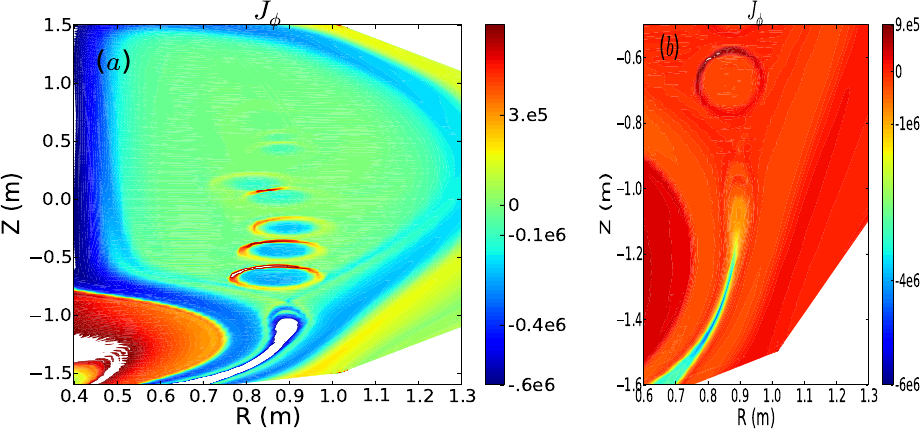}
\caption{Plasmoid formation in MHD simulations of NSTX-U with $\Psi_{\mathrm{inj}}=0.065Wb$ and strong poloidal flux shaping (t=7.93ms). Poloidal R-Z cut of toroidal currents (a) for the whole device  (b) zoom around the injector region. }
  \label{fig:fig5} 
\end{figure}

To further investigate the formation of plasmoids, we have also performed MHD simulations in the NSTX-U configuration. The location of poloidal flux coils are the same as in the experiment. However, the coils currents in these simulations are optimized to give a 
very narrow injector flux footprint width~\cite{ebrahimi2014}. This would give rise to a much more effective forced reconnection compared to NSTX simulations shown above. Due to the much larger reconnecting field (B=500G),  
the Lundquist number is as high as 29000 and well into the unstable regime in NSTX-U. As a result, compared to NSTX cases, plasmoid instability occurs during the injection phase (while the injector voltage is applied).  Plasmoid instability with  continued injection of 
plasmoids (up to eight) are observed as shown in Fig~\ref{fig:fig5}(a). The length of the unstable elongated current sheet shown in Fig~\ref{fig:fig5}(b) can reach up to 0.6m, which is shorter than that obtained above for NSTX. 

In summary,  in a large-scale toroidal fusion plasma, formation of 
plasmoids has for the first time been demonstrated by simulations, accompanied by promising camera data from experimental observations. Because (a) the high-$S$ plasma regime, which is very difficult to access in experiments designed for reconnection, can be accessed during CHI, and (b) there is no pre-existing plasma or
current channel in a transient CHI discharge, the system  
provides an ideal
laboratory for controlling and altering some of the parameters governing $S$
and the current sheet to better understand conditions under which small and
large plasmoids are generated. Understanding these plasmoids may be necessary to
predict how transient CHI scales as it is extrapolated to future (larger)
devices, such as the ST-FNSF (Fusion Nuclear Science Facility).
Plasmoid instability could be the leading mechanism for fast flux closure, and its control may be important
for the closed flux plasma stability as considerably larger amounts of poloidal flux are incorporated in
future devices to increase the start-up current magnitude. In addition, the two key characteristics of results presented here; (a) fast reconnection, and (b) formation of MHD global system-size plasmoid formation in the presence of guide field are relevant to solar observation of fast reconnecting plasmoids. These results along with future experiments with high-resolution diagnostics could improve our understanding of guide-field reconnection with strong implications for astrophysical plasmas.

\acknowledgements
Many thanks to Dr. N. Nishino of Hiroshima
University for the fast camera measurements. We also acknowledge Drs. H. Ji, A. Bhattacharjee, J. Menard, and S. Prager for useful discussions and comments on this Manuscript. This work was supported  by DOE-FG02-12ER55115 and CMSO. This work was also 
facilitated by the Max-Planck/Princeton Center for Plasma Physics.

\end{document}